\documentclass[12pt]{article}
\usepackage{amsmath}
\usepackage{amssymb}
\usepackage{amsthm}
\usepackage[titletoc,title]{appendix}
\usepackage{graphicx}
\numberwithin{equation}{section}
\usepackage[left=0.98in,right=0.98in,bottom=1.5in]{geometry}
\usepackage{setspace}
\usepackage [linktocpage]{hyperref}
\begin{document}
\title{\bf A Truncation of 11-Dimensional Supergravity \\ for Fubini-Like Instantons in AdS$_4$/CFT$_3$  \vspace{0.5cm} }

\author{{\bf M. Naghdi \footnote{E-Mail: m.naghdi@ilam.ac.ir} } \\
\textit{Department of Physics, Faculty of Basic Sciences}, \\
\textit{University of Ilam, Ilam, West of Iran.}}
\date{\today}
 \setlength{\topmargin}{-0.1in}
 \setlength{\textheight}{9.2in}
  \maketitle
  \vspace{-0.15in}
    \thispagestyle{empty}
 \begin{center}
   \textbf{Abstract}
 \end{center}
From a consistent Kaluza-Klein truncation of 11-dimensional supergravity over $AdS_4 \times CP^3 \ltimes S^1/Z_k$, with a general 4-form ansatz, we arrive at a set of equations and solutions for the included fields. In particular, we have a bulk equation for a self-interacting (pseudo)scalar with arbitrary mass. By computing the energy-momentum tensors of the associated Einstein equations, to include the backreaction, and setting them to zero, we solve the resulting equations with the bulk one and get solutions corresponding to marginal and marginally relevant deformations of the boundary CFT$_3$, which break the conformal symmetry. These bulk (pseudo)scalars are $SU(4) \times U(1)$-singlet and the corresponding solutions break all supersymmetries and parity because of the associated (anti)M-branes wrapping around specific and mixed internal and external directions. As a result and according to AdS$_4$/CFT$_3$ duality rules, we would realize the boundary counterpart in three-dimensional Chern-Simon $U(N)$ field theories with matters in fundamental representations. In particular, we build a $SO(4)$-invariant Fubini-like instanton solution by setting a specific boundary Lagrangian. The resulting solution is used to describe the dynamics of thin-wall bubbles that cause instability and big crunch singularities in the bulk because of the unboundedness of the boundary double-hump potential from below. Relations to mass-deformed ABJM model, $O(N)$ vector models and scale invariance breaking are also discussed. Meanwhile, we evaluate corrections for the background actions because of the bulk and boundary instantons.

\newpage
\setlength{\topmargin}{-0.7in}
\pagenumbering{arabic} 
\setcounter{page}{2} 


\section{Introduction}
In a few recent studies, we have considered Kaluza-Klein reductions of 11-dimensional (11D) supergravity (SUGRA) over $AdS_4 \times S^7/Z_k$ when the internal space $S^7/Z_k$ is an $U(1)$ bundle on $CP^3$. As a result, we found localized or partially localized objects in the bulk of AdS$_4$, and explored their boundary CFT$_3$ holographic duals according to the well-known AdS/CFT correspondence rules (see \cite{Aharony99} as an original review). Following the special truncation considered in \cite{Me7} in probe approximation, here we include also the backreaction so that the truncation would be consistent.

In fact, after considering a general 4-form ansatz of the 11D SUGRA and facing our setups with those in \cite{Gauntlett03}, where a more general 4-form ansatz and truncation are analyzed, we obtain the equations of motion (EOM’s) for the involved (pseudo)scalars in Euclideanized Anti-de Sitter space ($EAdS_4$) and get some solutions including the backreaction. Next, we take a specific 4-form ansatz and obtain a second-order nonlinear partial deferential equation for an included (pseudo)scalar that could be massive, massless or tachyonic and is self-interacting. To address the backreaction, we compute the energy-momentum (EM or stress) tensors of the Einstein equations because of the bulk flux turned on, and get a new set of scalar equations, in external and internal spaces, which must in turn be satisfied to insure that the resulting solution or object does not backreact on the background geometry. By solving them together with the main bulk equation, we see that to have instantons we have to take an exactly marginal or a marginally relevant deformation on the boundary \footnote{It is noteworthy that the exactly marginal solution, which is obtained by setting the bulk stress tensors to zero, might be attributed to a bulk (pseudo)scalar profile with $SO(4,1)$ symmetry--we have already analysed such massless modes in \cite{Me4}.}.

The resulting (pseudo)scalars are singlets of $SU(4) \times U(1) \equiv H$ (as the isometry group of the whole internal space) and, at least because of the associated (anti)M-branes wrapping around the mixed directions, break all supersymmetries (SUSY's) ($\mathcal{N}=6 \rightarrow 0$) according to the intersection rules of M-branes, for instance, in \cite{Bergshoeff96}. Meanwhile, we notice that a reasonable way to justify the SUSY breaking and get the needed singlet operators on the boundary theory, is to swap the fundamental representations (reps) of $SO(8)\equiv G$ (as the isometry group of $S^7$) for supercharges, fermions and scalars; and then discuss how we can get the wished $H$-singlets under the branching $G \rightarrow H$. In addition, the mass term in the bulk action breaks the scale invariance (SI) and that, although the bulk solution including the backreaction is scale-invariant in leading order, a relevant or mass term beside a marginal term in the boundary Lagrangian breaks the conformal $SO(4,1)$ symmetry.

Indeed, as a dual description for the bulk solution including the backreaction, we first consider a scalar Lagrangian with a marginal deformation term (or a triple-trace deformation of a relevant dimension-one operator) plus a so-called mass term that meet our needs well; see \cite{Elitzur05}. Next, we make instanton solutions for the massless case and argue that they should be $SO(4)$-invariant on $S^3$ (or $SO(3,1)$ on 3D de Sitter space-time $dS_3$ after the Lorenz analytical continuation), cause vacuum instability and are dual to big crunch singularities in the bulk. The solution's moduli $a$ and $\vec{u}_0$ mark size and location of the boundary instanton and of a thin-wall bubble that may nucleate everywhere in the bulk. The boundary duals might be realized in 3D Chern-Simon (CS) $U(N)$ and $O(N)$ vector models, although we continue to use the former model \footnote{The ABJM model \cite{ABJM} in large $k$ reduces to a 3D $O(N)$ vector model and on that basis, a marginal triple-trace deformation with a well-known Ultra-Violet (UV) fixed-point at $g_6^*=192$ is studied in \cite{CrapsHertog}; we return to this issue in subsection \ref{subsec.5.2}.}. Especially, we focus on an $U(1)$ part of the ABJM quiver gauge group with matters in fundamental reps of $SU(4)$, because of symmetry arguments and correspondence rules.

The organization of this paper is as follows: In section 2, we introduce the background and a general 4-form ansatz of 11D SUGRA. Then, we obtain its corresponding equations and solutions in $EAdS_4$ space, without and with including the backreaction, in Appendix A from facing our ansatz with that in \cite{Gauntlett03}. In section 3, we use a special version of the general ansatz, which results in an interesting bulk equation; and to get the solutions including the backreaction, we first compute the associated EM tensors of the Einstein equations in Appendix B (with some useful formulas used in computations in subappendix B.1) and then write the EOM's by zeroing both the external and internal components of the stress tensors. As a result, we obtain the main solution and conditions arisen from solving the latter equations and the main one in the bulk simultaneously in subsection 3.1. In subsection 3.2, we evaluate the correction to the background 11D action based on the solution including the backreaction. In section 4, we discuss the symmetries of the bulk setup and solutions and argue how they can help to fulfill the state/operator correspondence and find the correct boundary counterparts. Section 5 deals with the field theory duals of the main bulk solution including the backreaction. There, we present a suitable dual Lagrangian, find a plain solution and interpret implications of it. In particular, we discuss the relation of the latter setup to a massive deformation of the ABJM model and other issues such as SI breaking in subsection 5.2. In section 6 is a summary and more comments especially on the instability and false vacuum decay because of the instanton.

\section{The background and Genaral Ansatz} \label{sec02A}
The background we consider is 
\begin{equation}\label{eq01A}
 \begin{split}
 & ds^2_{11D}= {R_{AdS}^2}\, ds^2_{AdS_4} + R_7^2\, ds^2_{S^7/Z_k}, \\
 & G_4^{(0)}= d\mathcal{A}_3^{(0)} = 3\, R_{AdS}^3\, \mathcal{E}_4 = N \mathcal{E}_4,
 \end{split}
\end{equation}
where the geometry is $AdS_4 \times S^7/Z_k$ of 11D SUGRA with $S^7$ as a $S^1$ fiber-bundle on $CP^3$, and the 4-form ansatz is that in ABJM \cite{ABJM}. 
The general 4-form ansatz we are considering here is
\begin{equation}\label{eq01}
 \begin{split}
G_4 = & \left( 3 f_1 + R_{AdS} \ast_4(df_2 \wedge \mathcal{A}_3^{(0)}) \right) \wedge J^2 - R_{AdS}^{-1} \left( df_3 - f_4\, \ast_4 \mathcal{A}_3^{(0)} \right) \wedge J \wedge e_7  \\
& + \frac{1}{4 R_{AdS}^3} \ast_4(df_5 \wedge \ast_4 \mathcal{A}_3^{(0)}) \wedge J + \frac{3}{16 R_{AdS}^5} \ast_4 df_6 \wedge e_7 + \frac{3}{64 R_{AdS}^3}  f_7\, \mathcal{E}_4,
 \end{split}
\end{equation}
where $f_1, f_2, \ldots$ are scalar functions in the external space, $R=R_7= 2 R_{AdS}$ is the $AdS$'s radius of curvature, $\mathcal{E}_4$ is the unit-volume 4-form on $AdS_4$, the 2-form $J=d \omega$ is the K\"{a}hler form on $CP^3$, and $e_7 = (d\acute{\varphi}+\omega)$ with $\acute{\varphi}$ as the fiber coordinate. 

For the general 4-form anstaz (\ref{eq01}), we have derived the equations and conditions arising from satisfying the Bianchi identity ($dG_4=0$) and the Euclidean EOM
\begin{equation}\label{eq01B}
d \ast_{11} G_4 - \frac{i}{2}\, G_4 \wedge G_4=0
\end{equation}
in \cite{Me7}. However, it is also interesting to discuss the backreaction, that is considering the Einstein's equations as well; and of course we have done it in Appendix \ref{Appendix.AA}, where equations and solutions without and with including the backreaction are analysed in accordance with computations in \cite{Gauntlett03} where a more general ansatz is employed.

\section{Special 4-Form Ansatz and Solutions} \label{sec03}
We employ a special 4-form ansatz from (\ref{eq01}), made of the plain forms $e_7, J, \mathcal{E}_4$, whose clear form reads 
\begin{equation}\label{eq11}
\frac{\tilde{G}_4}{(2 R_{AdS})^4} = 8\, \bar{f}_1 J^2 - 2\, df_3 \wedge J \wedge e_7 + \frac{3}{8} f_7\, \mathcal{E}_4,
\end{equation}
where $f_1 N =\bar{f_1}$; and the resulting EOM becomes
\begin{equation}\label{eq12}
     \ast_4 d \left(\ast_4 d\bar{f}_1 \right) - \frac{4}{R^2} \left(1 \pm 3\, \bar{C}_7 \right) \bar{f}_1 - 2 \times 192\, \bar{f}_1^3=0,
\end{equation}
where for different values of $\bar{C}_7$, towers of tachyonic, massless (with $\bar{C}_7 =\frac{1}{3}$) and massive (pseudo)scalars in the bulk of $EAdS_4$ are possible.

It is interesting to compare the ansatz (\ref{eq11}) with (2.5) in \cite{Gauntlett03}. As a result, we see that with $U = V = \chi = A_1 = B_1 = B_2 = 0$ and
\begin{equation}\label{eq13a}
     f = 6\, f_7, \quad h= 4\, R^4 \bar{f}_1, \quad dh= - R^4\, df_3,
\end{equation}
the formalisms match, and the counterpart of (2.23) in \cite{Me7} is
\begin{equation}\label{eq13b}
     f = \frac{6}{R^7} \left(\epsilon + h^2 \right), \quad \epsilon =\pm\, \bar{C}_7\, R^6,
\end{equation}
which comes from the equation (B.11) of \cite{Gauntlett03}. In particular, we read from (B.13)
\begin{equation}\label{eq13c}
     \ast_4 d (\ast_4 dh)- (16 + 24 \epsilon)\, h - 2 \times 12\, h^3 =0,
\end{equation}
which is the same as  (\ref{eq12}) up to some scaling and notice that $R=1$ is set. Then, from the latter equation, we read $m^2 R_{AdS}^2=-2$ with $\epsilon = -1$ (skew-whiffed) and $m^2 R_{AdS}^2=10$ with $\epsilon = 1$ (Wick-rotated) \footnote{Remember that with $\eta = 2 e_7$, the modes in (\ref{eq13c}) match with those in (\ref{eq12}).}, which were already discussed in \cite{Me6} and \cite{Me7} respectively. \\
To continue, we note that considering only the EOM (\ref{eq12}) means working in probe approximation, that is excluding the backreaction, for which we presented some solutions in \cite{Me7} with dual descriptions. Here we include the backreaction as well.

\subsection{Solutions Including the Backreaction} \label{subsec03.01}
To get solutions including the backreaction, we should first compute the stress-energy tensors of the replying Einstein equations. The details of such computations are given in Appendix \ref{Appendix.A0}, where (\ref{eq20a}), (\ref{eq20b}) and (\ref{eq20c}) come from zeroing the external, internal and seventh components of the EM tensors, respectively. Next, by combining the latter equations with the main one (\ref{eq12}), we arrive at
\begin{equation}\label{eq23a}
     \Box_4 \bar{f}_1 =0,
\end{equation}
\begin{equation}\label{eq23b}
     \Box_4 \bar{f}_1 + \left(8 \pm 12\, \bar{C}_7 \right) \frac{\bar{f}_1}{R^2}=0,
\end{equation}
\begin{equation}\label{eq23c}
     \Box_4 \bar{f}_1 + \frac{8}{3 R^2} \bar{f}_1=0,
\end{equation}
respectively, which are conditions imposed on the (pseudo)scalar from including the backreaction on the background geometry of the external and internal spaces. In other words, if we find solutions to these equations, the corresponding objects (e.g. instantons as topological objects) do not backreact on the background geometry.

However, satisfying (\ref{eq23a}), (\ref{eq23b}) and (\ref{eq23c}) simultaneously results in $\bar{f}_1=0$; Still, we may discuss each of them separately. In particular, the solution
\begin{equation}\label{eq24a}
      \bar{f}_1(u,\vec{u}) = c_{8} + \frac{c_{9}\, u^3}{\left[u^2 + (\vec{u}-\vec{u}_0)^2 \right]^3}
\end{equation} 
of (\ref{eq23a}), which corresponds to a \emph{marginal} deformation with the boundary operator $\Delta_+ =3$, does not backreact on the external space geometry.

On the other hand, satisfying (\ref{eq23b}) and (\ref{eq23c}) simultaneously, which means taking the backreaction in the whole internal space into account, again results in the trivial solution $\bar{f}_1=0$; but if we take the massless solution (\ref{eq24a}), which is in turn realized with $\bar{C}_7 = \frac{1}{3}$ (given that $(1 \pm 3\, \bar{C}_7)={m^2} {R_{AdS}^2}$) in the skew-whiffed version of (\ref{eq12}), in the internal $CP^3$ space equation (\ref{eq23b}), we will have
\begin{equation}\label{eq24b}
 \Delta_{\pm} = \frac{3}{2} \pm \frac{\sqrt{5}}{2},
\end{equation}
which corresponds to a \emph{marginally relevant} operator with $\Delta_+ < 3$ (and the same behavior for \ref{eq23c}); That means for this solution does not have any backreaction on the background metric, one must take a marginally relevant deformation in the corresponding boundary theory; We return to this interesting case when discussing dual solutions in section \ref{sec.5}. \footnote{It is also interesting to consider solutions for each of the equations (\ref{eq23a}), (\ref{eq23b}) and (\ref{eq23c}) separately, in the same way done in subsection (2.4) of \cite{Me7} for the equation (\ref{eq12}).}

\subsection{Correction to the Action} \label{subsec03.02}
In is also interesting to compute the 11D action correction based on the solution including the backreaction. To this end, we use the 11D SUGRA action in Euclidean space as
\begin{equation}\label{eq25}
  S_{11D}^E = -\frac{1}{2 \kappa_{11}^2} \left[ \int d^{11}x \, \sqrt{g} \, \mathcal{R}_{11} + \frac{1}{2} \int \left(\tilde{G}_4 \wedge \ast_{11} \tilde{G}_4 - \frac{i}{3} \tilde{\mathcal{A}}_3 \wedge \tilde{G}_4 \wedge \tilde{G}_4 \right) \right],
\end{equation}
where $\mathcal{R}_{11}$ is the Ricci scalar and $\kappa_{11}^2 = 8\pi \mathcal{G}_{11}$ with $\kappa_{11}$ as the gravitational constant. To evaluate the correction, we should employ (\ref{eq11}) for $\tilde{G}_4$ and its 11D dual as
\begin{equation}\label{eq26a}
  \frac{\tilde{G}_7}{(2 R_{AdS})^7} = \bar{f}_1\, \mathcal{E}_4 \wedge J \wedge e_7 + \ast_4 df_3 \wedge
  J^2 +  f_7\, J^3 \wedge e_7,
\end{equation}
and that
\begin{equation}\label{eq26b}
\tilde{G}_4 = d\tilde{\mathcal{A}}_3, \quad \tilde{\mathcal{A}}_3 = \tilde{\mathcal{A}}_3^{(0)} + (8\, R^8) (\bar{f}_1\, J \wedge e_7), \quad \tilde{G}_4^{(0)} = d\tilde{\mathcal{A}}_3^{(0)} = \frac{3}{8} R^4 f_7\, \mathcal{E}_4.
\end{equation}
Next, plug (\ref{eq26a}) and (\ref{eq26b}) into (\ref{eq25}) together with
\begin{equation}\label{eq27}
 df_3 = -4\, d\bar{f}_1, \quad  f_7 = +i\, 32\, R\, \bar{f}_1^2 \pm i\, \frac{\bar{C}_7}{R},
\end{equation}
which in turn come from (\ref{eq13a}) and (\ref{eq13b}), the right part of the action for us (the second and third terms) becomes $\bar{S}_{11D}^E = S_0 + S_{11}^{modi.}$, where
\begin{equation}\label{eq28a}
  S_0 = \frac{9}{R^2\, \kappa_{11}^2} \texttt{vol}_4 \wedge \texttt{vol}_7, \quad \texttt{vol}_7 = \frac{R^7}{3!} \int  J^3 \wedge e_7 = \frac{\pi^4\, R^7}{3\, k},
\end{equation}
is from the ABJM background realized with $\bar{C}_7=1$, and
\begin{equation}\label{eq28b}
  S_{11}^{modi.} = \frac{3\, R^4}{2 \kappa_{11}^2}\, \texttt{vol}_7 \int \bigg(8\, R^2\, \bar{f}_1^2\, \mathcal{E}_4 + 32\, d\bar{f}_1 \wedge \ast_4 d\bar{f}_1 + 384\, R^2\, \bar{f}_1^4\,  \mathcal{E}_4 \bigg).
\end{equation}
Then, by putting the solution (\ref{eq24a}) with $c_8=0$ in the latter action and taking (see \cite{Me6})
\begin{equation}\label{eq29}
  \mathcal{E}_4 = \frac{1}{u^4}\, dx \wedge dy \wedge dz \wedge du, \quad \kappa_{11}^2 = \frac{16\, \pi^5}{3} \sqrt{\frac{R^9}{3\, k^3}},
\end{equation}
we arrive at, the correction in the unit 7D internal volume,
\begin{equation}\label{eq28c}
  S_{corr.} = \frac{9\, c_9^2}{32\, \pi^3} \sqrt{3\, k^3\, R^3} \left[\frac{35}{48} \frac{1}{\epsilon^6} + \frac{4199}{8192} \frac{c_9^2}{\epsilon^{12}}\right],
\end{equation}
as the finite part of the action, where $\epsilon>0$ is a cutoff parameter used instead of $u=0$ to evade the infinity of integrals with respect to (wrt) $u$.

\section{Dual Symmetries and Correspondence} \label{sec.04}
We first remind that the ansatz (\ref{eq11}) and the (pseudo)scalars therein are singlets of $SU(4) \times U(1)$ and so, we look if we can find the wished singlet (pseudo)scalars in the spectrum of the involved 11D SUGRA over $AdS_4 \times CP^3 \ltimes S^1/Z_k$. This task was already done in \cite{Me7}, where we considered three massive (pseudo)scalars. But, for the solution including the backreaction, the (pseudo)scalars should be almost massless and so, we should look whether we can find any singlet massless (pseudo)scalar in the spectrum or not.

To this end, we first note that the massless multiplet ($n=0$) of $G$ (as the isometry group of $S^7$) includes a graviton ($\textbf{1}$), a gravitino ($\textbf{8}_s$), 28 spin-1 fields ($\textbf{28}$), 56 spin-$\frac{1}{2}$ fields ($\textbf{56}_{s}$), 35 scalars ($\textbf{35}_{v}$) of $0_1^+$ emerging from the external ingredients ($\mathcal{A}_{\mu \nu \rho}$), and 35-pseudoscalars ($\textbf{35}_{c}$) of $0_1^-$ emerging from the internal ingredients ($\mathcal{A}_{m n p}$). In the higher KK multiplets ($n>0$), the massless pseudoscalar sets in $\acute{\textbf{840}}_{s}$ of $0_1^-$ with $n=2$ while the massless scalar sets in $\textbf{1386}_{v}$ of $0_1^+$ with $n=4$ of $G$ (see, for instance, \cite{FreedmanNicolai}, \cite{NilssonPope} and  \cite{Biran}), and there is not any $H$-singlet under the branching $G\rightarrow H$.

On the other hand, we read from the ansatz structure that it breaks all SUSY's because of the mixed internal directions around which the associated (anti)M-branes wrap-- see also \cite{DuffNilssonPope84} that states the solutions with 4-form components all in the internal space break SUSY's and parity; and as a result, the boundary duals might be realized in CS-matter $U(N)$ and $O(N)$ vector models. Thus, a consistent way to meet this need is to swap the fundamental reps $\textbf{8}_s$, $\textbf{8}_c$ and $\textbf{8}_v$ of $SO(8)$. On it, after swapping $\textbf{s} \leftrightarrow \textbf{v}$ with $\textbf{c}$ rep fixed (that means exchanging supercharges(spinors) with scalars while keeping the fermions unchanged), we have the rep $\textbf{1386}_{s}$ whose $U(1)$-neutral reps under the branching read
\begin{equation}\label{eq30}
      \textbf{1386}_{s} \rightarrow \textbf{1}_{0}\oplus \acute{\textbf{20}}_{0} \oplus \textbf{105}_{0} \oplus \textbf{336}_{0} \oplus ...\ ,
\end{equation} 
which include an $H$-singlet mode. On the other hand, for the massless pseudoscalar of the original model ($\acute{\textbf{840}}_{s}$), there is not any singlet under $H$ even after both swappings.

As another point, we note that the ansatz breaks the inversion symmetry and thus conformal transformation of $SO(4,1)$ (as isometry of $EAdS_4$) besides the fact that the mass term in the bulk equation (\ref{eq12}) breaks the SI; and as a result, we argue that our solution must be $SO(4)$ invariant; see \cite{Me7} for more details. It is also notable that although the resulting equation (\ref{eq23a}) and main solution (\ref{eq24a}) preserve full conformal symmetry, the marginally relevant deformation breaks the SI as we will discuss in the next section.

\section{Boundary Field Theory Duals} \label{sec.5}
For the general ansatz (\ref{eq01}), the scalar profiles of the forms (\ref{eq03a}) and (\ref{eq03d2}) are already discussed in \cite{Me4} and \cite{N} respectively and so, we do not pay more attention to them. Instead, we focus on the duals for the solutions including the backreaction. \\
Indeed, from the bulk description with backreaction in subsection \ref{subsec03.01}, we see that the dual boundary operator have to be for an exactly marginal or a marginally relevant deformation. Although in \cite{Me4}, \cite{Me5} and \cite{Me3} we studied samples of marginal operators and deformations, here we concentrate on a special sample of the (exactly) marginal deformation $\Delta_+=3$, valid as an approximation for another case too, and study aspects of it. In our formalism, we make this operator from the singlet (pseudo)scalar after the swapping $\textbf{8}_s \leftrightarrow \textbf{8}_v$ of the original supersymmetric theory. On the other hand, besides breaking SUSY's, the $H$-singlet states break the even-parity symmetry of the ABJM model, which might in turn be understood through fractional M2-brane \cite{ABJ} associated with the probe (anti)M5-brane wrapping around $R^3 \times S^3/Z_k$. As a result and after gauging, we remain with just the diagonal $U(1)$ part of the quiver gauge group $SU(N)_k \times SU(M)_{-k}$, for which we set $A_i^-=0$, as it is for the boundary baryonic symmetry under which our modes are singlet \footnote{It should be noted that the singlet sections of $U(N)$ and $O(N)$ CS vector models, with complex and real (pseudo)scalars in fundamental reps, are dual to the non-minimal and minimal Vasiliev higher-spin (HS) bulk theories (see \cite{Giombi01} for a review), respectively, with parity breaking scheme \cite{Aharony1110}, \cite{Minwalla1207}. As a result, our setups may be recast in that framework; For instance, our bulk massive modes may be considered as arisen from some loop corrections or interactions in the minimal HS model or as fluctuations about the minimal solution of $m^2 R_{AdS}^2=-2$.}. In other words, the fundamental fields are neutral to the diagonal $U(1)$ that couples to $A_i^+ \equiv (A_i  + \hat{A}_i)$ while $A_i^-$ acts as the baryonic symmetry and, since our (pseudo)scalars are neutral, we assign zero to it.

By the way, we could consider the marginal deformation as a triple-trace deformation \footnote{It is notable that with mixed boundary conditions, corresponding to multi-trace deformations, the effective action reads ${\Gamma}_{eff.} [\alpha] = S_{on}[\alpha] + \int d^3 \vec{u}\ \tilde{f}(\alpha)$, where $S_{on}$ is the bulk on-shell action and $\tilde{f}(\alpha)$ is a function of the local operator $\mathcal{{O}}_{\Delta_-}$ with which the action is deformed. In general, the mixed boundary conditions lead to conformal field theories only if $\tilde{f}(\alpha) \sim \alpha^{3/\Delta_-}$ or $\beta = f_0\, \alpha^{({3}/{\Delta_-})-1}$, where $\alpha$ and $\beta$ act as vacuum expectation value and source for the operator $\Delta_-$ and conversely for $\Delta_+$, and different values of $f_0$ correspond to various points along the lines of marginal deformations. \label{ftn.10.}} of the operator $\mathcal{O}_1 \sim \texttt{tr}(y \bar{y})$ used in \cite{Me5}. The most recognized case with the latter operator is the $O(N)$ vector models made of it and its multi-trace deformations. The interesting case is the tri-critical model, which includes just the kinetic term and a term proportional to a triple-trace deformation of it; see for instance \cite{Elitzur05}. With respect to the bulk studies, as a reasonable proposal consistent with our discussions, we match the bulk to boundary field as $\bar{f}_1 \rightarrow \varphi^2$ that in turn means the bulk instantons are square of the boundary ones \cite{deharo06}, in leading order of course. As a result, from the EOM (\ref{eq12}), we can take a dual Lagrangian for the boundary theory. Fortunately, a form for such an effective Lagrangian is already known (see \cite{Elitzur05} and \cite{deHaro2}) and follows our wishes. In fact, next to the CS term (here we continue to use the $U(N)$ model), we can consider \footnote{Note that we use the metric signature $(-, +, +,\ldots +)$ for both gravity and field theories; and after Wick-rotation, we reach to the fully positive signature metric along with $t_M \rightarrow i t_E$ and $e^{-i S_M} \rightarrow e^{-S_E}$, where $S_E$ is the positive Euclidean action.}
\begin{equation}\label{eq31}
     \mathcal{L}_{3}^{eff.} = \frac{1}{2} (\partial_i \varphi)^2+ \frac{1}{16} \mathcal{R}_3\, \varphi^2 - \frac{\lambda_6}{6 N^2}\, \varphi^6,
\end{equation}
where $\mathcal{R}_3$ is the boundary 3D scalar curvature that is $\frac{6}{R_0^2}$ for the three-sphere ($S^3$) of the radius $R_0$ \footnote{In fact, wrt the footnote \ref{ftn.10.} and fact that the solution (\ref{eq24a}) could be interpreted as a marginal triple-trace deformation, we can read the holographic effective action $\tilde{\Gamma}_{eff.} [\alpha]$, from the bulk analysis as shown in \cite{deHaro2}, which agrees with (\ref{eq31}).}. Such a Lagrangian is actually used (with $\lambda_6 > 0$) as a dual to describe the dynamics of Coleman-de Luccia large-expanding vacuum bubbles of $AdS_4$ in thin-wall approximation; see \cite{Barbon1003} and \cite{Maldacena010} for related discussions. In our setup, these shells might emerge from the special (anti)M5-brane wrapping that results in domain-walls interpolating among the bulk vacua; see also \cite{Bena}. In other words, note that with $\lambda_6 > 0$, we have an unbounded potential from below signaling instability near the potential extrema, which in turn describes the bulk big crunch singularities. Indeed, the $O(3,1)$ invariant bubble solution includes an open and infinite Friedmann-Lema\^{\i}tre-Robertson-Walker universe inside $AdS_4$ space-time that collapses to a big crunch singularity. A field theory dual for the latter is obtained from a marginal triple-trace deformation of the ABJM model in \cite{CrapsHertog}; see also \cite{BzowskiHertog} for a recent related study.

\subsection{An Explicit SO(4)-Invariant Solution} \label{subsec.5.1}
Because the classical solutions for the EOM from the Lagrangian with different $\mathcal{R}_3$'s could be related by conformal transformations, to have a simple analytical solution, we consider a massless instanton on $S^3_\infty$(the three-sphere at infinity, $R_0\rightarrow \infty$) that is in turn equivalent to the solution on $R^3$ and so, we set the second term of (\ref{eq31}) to zero for now and provide complementary discussions in the next subsection \footnote{An analysis with a Lagrangian like (\ref{eq31}), when the boundary is defined on $S^3$, is presented in \cite{SmolkinTurok}.}. Equally and to comply with our formalism of using the ABJM boundary action, we set the fermions to zero with keeping only the $U(1)_{diag}$ part of the gauge group beside the scalar part; That is
\begin{equation}\label{eq32}
  \mathcal{\bar{L}}_{3} = \mathcal{L}_{CS}^+  - \texttt{tr} (\partial_i Y_A^{\dagger}\, \partial^i Y^A ) - V_{bos},
\end{equation}
where $\mathcal{L}_{CS}^+$ is the CS Lagrangian associated with $A_i^+$ and $V_{bos}$ is the bosonic potential of ABJM; see, for instance, \cite{Me4}. Then, to make a proper solution, we take the following ansatz
 \begin{equation}\label{eq33}
     Y^A = i\, \varphi(r)\, S^A, \quad  Y_A^\dagger = \varphi(r)\, S_A^\dagger, \quad S^A= S^B S_B^\dagger S^A - S^A S_B^\dagger S^B,
\end{equation}
where the presence of $i$ factor and using different $Y^A$ and $Y_A^\dagger$ are because of being in Euclidean space; and further set $Y^3=Y^4=0$. A clear solution for $S^1$ and $S^2$ reads \cite{Gomiz}, \cite{Terashima}
\begin{equation}\label{eq33a}
    (S_1^\dagger)_{m,n} = \sqrt{m-1}\, \delta_{m,n}, \quad (S_2^\dagger)_{m,n} = \sqrt{N-m}\, \delta_{m+1,n}.
\end{equation}
Thereupon, the scalar EOM of $Y_A^\dagger$ becomes
\begin{equation}\label{eq34}
    \partial_i \partial^i \varphi(r) + \frac{12 \pi^2}{k^2}\, \varphi(r)^5 =0,
\end{equation}
from which a solution reads
\begin{equation}\label{eq34a}
   \varphi(r)= \sqrt{\frac{k}{2 \pi}}\, \left( \frac{a}{a^2 + (\vec{u}-\vec{u}_0)^2} \right)^{1/2},
\end{equation}
which is Fubini-like \cite{Fubini1} and the $O(4)$-invariant one we have been looking for. Meanwhile, we notice that by the ansatz (\ref{eq33}), the gauge field equations are satisfied trivially. As a commentary, note that because of the SI in its UV fixed-point, the model admits an infinite family of instantons responsible to form the bulk vacuum bubbles, which in turn break the $AdS_4$ isometry to $SO(3,1)$. The broken generators, with four free parameters $a$ and $\vec{u}_0$ marking the size and position of the boundary instanton respectively, act to translate the bubbles around in the bulk 4D volume. Further, the finite contribution for the action on $S^3_\infty$ based on the solution (\ref{eq34a}), after linearization of $S^A$ matrices, reads
\begin{equation}\label{eq34b}
   \int_0^\infty \frac{r^2}{(a^2 + r^2)^2}\, dr = \frac{\pi}{4\, a} \ \ \Rightarrow \ \ S_{b} \cong -\frac{\pi}{2 k\, \lambda^2},
\end{equation}
where in the last term on the RHS we have used the large $N$ normalization coefficient $1/N^4$ for the marginal deformation term and $\lambda= N/k$.

Still, a subtle point is that as our bulk solutions are $H$-singlets, how we can use the ansatz (\ref{eq33}). To resolve this, we first note that $Y^A$ and $Y_A^\dagger$ may be considered as independent degrees of freedom in Euclidean space, which signals the parity breaking as well. Next, we notice that with $Y^3=Y^4=0$, one of the $SU(2)$'s is considered as a spectator and then, from one of the two complex scalar fields of the remaining $SU(2)$, we may write
 \begin{equation}\label{eq35}
     Y \equiv Y_1^\dagger + Y^1, \quad  Y^\dagger \equiv Y_2^\dagger + Y^2;
\end{equation}
Last, wrt the swapping $\textbf{s} \leftrightarrow \textbf{v}$ and with a linear combination of these, we can build the singlet scalars as
 \begin{equation}\label{eq35a}
     y = \varphi(r) (S_1^\dagger S^2 - S^1 S_2^\dagger), \quad \bar{y} = i \varphi(r) (S^1 S_2^\dagger - S_1^\dagger S^2).
\end{equation}

\subsection{Relation to Mass Deformation and SI Breaking} \label{subsec.5.2}
An ansatz like (\ref{eq33}) is already used as a tool to make the fuzzy $S^3$ solutions of the mass-deformed ABJM model; see \cite{Gomiz}, \cite{Terashima} and \cite{HanakiLin}. In fact, the mass deformation of the original ABJM model, associated with turning the bulk flux on, is done through a relevant operator, $\mathcal{L}_{mass}=\mu^2 \texttt{tr}(Y^A Y_A^\dagger)$, which breaks the global R-symmetry $SU(4)$ down into $SU(2) \times SU(2) \times U(1)$ while breaks the conformal symmetry $SO(3,2)$ entirely. The vacua of the mass-deformed model have interpretations as fuzzy three-spheres. In other words, the M5-branes are understood as M2-branes puffing into a fuzzy sphere near the M5-brane core. The fuzzy-funnels solutions describing M2-M5 brane intersections have interpretations in the field theory on M2-branes as domain-walls. It is interesting that the second term of the Lagrangian (\ref{eq31}) has a similar structure to this mass deformation term.

There also are other interesting points and discussions with the Lagrangian and our setup. First note that the potentials of $O(N)$ vector models in three dimensions are renormalizable in the large $N$ limit; and by including relevant and marginal terms up to the so-called $\varphi^6$, the potentials are stable for $0\leq g_6 \leq g_6^c =(4 \pi)^2$ and unstable for $g_6 > g_6^c$ and $g_6<0$ because of the potential unbounded from below--note that $g_6 \equiv -\lambda_6$ and that the sign of $g_6$ controls the behavior of the system in the classical case. There is the tri-critical model just with $g_6 (\varphi^2)^3$ for which the beta function is zero at $N\rightarrow \infty$, while $1/N$ corrections break the SI and a massless Goldstone boson (dilaton) is appeared as a dynamical bound-state of $(\vec{\varphi}.\vec{\varphi})$--we remind that for any finite and positive $g_6$, the operator becomes irrelevant quantum mechanically. In the next-to-leading order of the $1/N$ expansion, the dilaton gets a tachyonic mass and the spontaneously broken phase becomes unstable. In other words, in infinite $N$ limit, there is the UV fixed-point $g_6^c \cong 158$ and by increasing $g_6$, one reaches to the UV fixed-point $g_6^* = 192$, which in turn is an instability region where non-perturbative effects are dominated \cite{Bardeen1984}; see also \cite{Elitzur05}. As a result, the nonzero vacuum expectation value $\langle \varphi^2 \rangle \neq 0$ is equivalent to the massive deformation discussed above or a relevant deformation by the $\Delta_{-}=1$ operator. In addition, this marginal plus relevant deformations are consistent with the marginally relevant operator we have got in (\ref{eq24b}) by including the backreaction.

Further, according to the discussions in \cite{RabinoviciSmolkin011}, in the presence of CS term, only the dilaton causes a bounded potential and for $g_6 > g_6^c$ just the neutral states under $U(1)$ are formed; and this is the unstable phase we have met as well. Still, a more interesting discussion is in \cite{BardeenMoshe014}, where spontaneous breaking of SI in 3D $U(N)_k$ CS theories coupled to a scalar in fundamental rep is studied in the large $N$ limit. By adding a self-interacting term like $\lambda_6 (y^\dagger y)^3$, they have shown that there is a massive phase for a critical combination of $\lambda=N/k$ (the t'Hooft coupling) and $\lambda_6$; Indeed, for the tri-critical model, there is a spontaneous SI breaking for $\lambda^2 + \frac{\lambda_6}{8 \pi^2}=4$. Near the critical phase, the $U(N)$-singlet massive bound-state plays role as a pseudo-dilaton. Meanwhile, this phase of the SI breaking is dual to the parity broken version of the Vasiliev's HS theory in $AdS_4$ bulk \cite{Vasiliev01}.

\section{Final Comments}
In this study, we first considered a general 4-form ansatz of 11D supergravity over $AdS_4 \times CP^3 \ltimes S^1/Z_k$ and after comparing its structure with that in \cite{Gauntlett03}, analyzed the resulting equations and solutions, especially by including the backreaction. Next, making use of a special 4-form ansatz with the same settings, we obtained a particular nonlinear EOM, from the identity and equation of the 4-form, that included all massive, massless and tachyonic modes for an involved (pseudo)scalar in $EAdS_4$. Then, we referred to the Einstein equations and got the equations (\ref{eq20a}), (\ref{eq20b}), (\ref{eq20c}) from setting the bulk energy-momentum tensors to zero. Then, by solving them simultaneously with the bulk equation (\ref{eq12}), we obtained an instanton solution in subsection \ref{subsec03.01}, from the consistent truncation, corresponding to an exactly marginal or a marginally relevant deformation of the boundary theory. After that, we evaluated a correction to the main background supergravity action in subsection \ref{subsec03.02}. After an analysis of dual symmetries, we finally discussed the corresponding field theory counterparts, solutions and some related points.

In fact, for the boundary dual of the bulk solution with considering the backreaction, we employed a scalar Lagrangian including a marginal deformation term, which could be considered as a triple-trace deformation of a dimension-one operator, plus a relevant mass term of the latter operator, like those used in the standard $O(N)$ and $U(N)$ vector models. Then, considering the correspondence rules, we built a $SO(4)$-invariant Fubini-like instanton solution. The condition to break scale invariance, except by a boundary mass term, bulk interpretations of the solutions and other related issues were also addressed. In particular, we noticed that the marginal deformation in the CFT$_+$ fixed-point triggers an instability of CFT$_3$ on $S^3$ (or $dS_3$ in Lorentzian signature) \footnote{Note also that multi-trace deformations in general destabilize $AdS_4$ vacua, see \cite{HertogMaeda01}, although the triple-trace deformation here preserves the conformal invariance in leading order.}.

As the last comment, we notice that because of the special (anti)M-brane wrapping, there are domain-wall flows corresponding to thin-wall bubbles of $AdS_-$ (for the true vacuum) that expand exponentially within $AdS_+$ (for the false vacuum) and show Bose-Einstein condensations on $dS_3$ space-time. In the case that the condenses lead to the bulk crunches, the instability dynamics is described by finite $N$ corrections including formation and collisions of multi-bubbles; see \cite{Elitzur05} and \cite{Barbon1003}. In other words, a $AdS_-$ thin-wall bubble expanding within $AdS_+$ is equivalent to a flow between the UV (CFT$_+$) and IR (CFT$_-$) fixed points. Arriving the bubbles in a finite time to the boundary has a CFT interpretation as rolling in the potential of the unstable marginal boundary operator. Meantime, it is argued in \cite{BarbonRabinovici011} that the crunches are associated with \emph{negative energy falls} at least for marginally relevant operators. One should also note that although the constant-field arrangements $\bar{\varphi}=0$ and $\bar{\varphi} = \pm \left[{3}/{(4 R_0^2\, \lambda)} \right]^{1/4}$ are the local minimum and maximums of the potential $V(\varphi)=\frac{3}{8\, R_0^2}\, \varphi^2 - \frac{\lambda_6}{6 N^2}\, \varphi^6$  in (\ref{eq31}) respectively, and have \emph{bounce} nature, the bubble dynamic is however described by a field similar to $\varphi(r)$ in (\ref{eq34a}). For discussions on \emph{generalized Fubini instantons} and \emph{oscillating Fubini instantons} of double-hump potentials of this type, see \cite{BumHoonLee014A} and \cite{BumHoonLee014}. \\

\begin{appendices}

\section{Solutions from Matching with \cite{Gauntlett03}} \label{Appendix.AA} 
In this Appendix, we discuss the equations, solutions and backreaction issue for the general ansatz (\ref{eq01}) briefly. To this end, we first face the formalism and reduction here for the metric and 4-form with $ds^2$ and $G_4$ given, respectively, in the equations (2.4) and (2.5) of \cite{Gauntlett03} \footnote{Note that our ansatz (\ref{eq01}) is a general one that could be constructed from the scalar functions in the external space next to the given forms $\omega, e_7$ and $\mathcal{A}_3^{(0)}$ of the background solution, while $G_4$ in \cite{Gauntlett03} includes more ingredients. Except for this overall likeness, we have our own objectives and to succeed them, we study the bulk modes, equations, solutions and other related topics as well as their field theory duals in details, none of which has been covered in \cite{Gauntlett03}.} that read
\begin{equation}\label{eq02a}
ds_{11}^2 = ds_4^2 + e^{2 U} ds^2 ({K E_6}) + e^{2 V} \left( \eta + A_1 \right)^2,
\end{equation}
 where $ds_4^2$ and $ds^2 ({K E_6})$ (a K\"{a}hler-Einstein metric) are equivalent to our full $AdS_4$ and $CP^3$ metric respectively, $U, V$ are scalar fields and $A_1$ is an 1-form on the external 4D space, and
\begin{equation}\label{eq02b}
 \begin{split}
 G_4^{(1)} = & 2 h\, J \wedge J + H_1 \wedge J \wedge  \left( \eta + A_1 \right) + H_2 \wedge J + H_3 \wedge \left( \eta + A_1 \right) + f\, \texttt{vol}_4 \\
   & + \sqrt{3}\, \left(\chi_1 \wedge \Omega + \chi\, \left( \eta + A_1 \right) \wedge \Omega + c.c. \right),
 \end{split}
\end{equation}
where $h, f$ are real scalars, $H_r\, (r=1,2,3,)$ are $r$-forms, $\chi_1$ is a complex 1-form and $\chi$ is a complex scalar all on the external 4D space, and $\Omega$ is the complex $(3,0)$ form on $CP^3$. 

To adjust the formalisms, we should first take $e_7 \rightarrow 2 e_7 \equiv \eta$  and as a result $J \rightarrow 2J$, and set $U = V = \chi_1 = \chi = A_1 = B_1=0$. Next, from comparing the ansatzs and satisfying the Bianchi identity $d G_4=0$, we have
\begin{equation}\label{eq03a}
 H_1=dh=0, \quad f_1= f_2= c_1 + c_2\, u^3, \quad h=\frac{3}{2} c_1, \quad df_3 = f_4\, \ast_4 \mathcal{A}_3^{(0)},
\end{equation}
where $c_1, c_2, \ldots$  are constants of integration, $u$ is the horizon coordinate in the Poincar$\acute{e}$ metric
\begin{equation}\label{eq03aa}
 ds^2_{EAdS_4} = \frac{1}{u^2} \left(du^2 + dx_i ^2 \right),
\end{equation}
with $i=(1,2,3)$ for $(x,y,z)=\vec{u}$ respectively, and
\begin{equation}\label{eq03b}
 f= \pm \frac{6}{R^7} \epsilon, \quad  \texttt{vol}_4 =\frac{R^4}{16}  \mathcal{E}_4,
\end{equation}
where $\epsilon = f_7 = - \frac{i}{3} R^3$, with $i$ for being in Euclidean space \footnote{One should note that with $G_4 \rightarrow i G_4$, the Euclidean equation (\ref{eq01B}) goes to that in \cite{Gauntlett03}; and that with $\epsilon =1$ in \cite{Gauntlett0912}, the matching term is just the ABJM background solution $G_4^{(0)}$.} and the lower sign for the skew-whiffed solutions in general, and
\begin{equation}\label{eq03c}
H_2 = \frac{2}{R^3} \ast_4(df_5 \wedge \ast_4 \mathcal{A}_3^{(0)}), \quad H_3 = \frac{3}{R^5} \ast_4 df_6,
\end{equation}
where $H_2 = 2 B_2$, $H_3 = dB_2$ \footnote{It is remarkable that, according to the discussion after the equation (2.18) in \cite{Gauntlett03}, the massive ($m^2 R_{AdS}^2=12$) 2-form $B_2$ could be dualised to a pseudoscalar ($a$) with $\Delta_+ = 5$, which might in turn be identified with $f_5$ we have considered in \cite{Me7}.}. Note also that to satisfy the Bianchi identity $dH_3 =0$ and $dH_2 =2 H_3$, the conditions read \footnote{Note that if we set $f_5$=0 in the main ansatz (\ref{eq01}), we have a massless (pseudo)scalar in $AdS_4$ that is already studied in \cite{Me4}.}
 \begin{equation}\label{eq03d}
 f_5=-\frac{3}{2 R} f_6, \quad \partial_i \partial^i f_5 = 0
\end{equation}
with $r=\sqrt{x_i x^i}$ and the solution
 \begin{equation}\label{eq03d2}
  f_5(r)=c_3 + \frac{c_4}{r}.
\end{equation}
It is also notable that to satisfy the EOM of $G_4$, an extra condition is setting $c_1=0$.

Now, we try to present solutions including the backreaction. To do this, from the action (2.10) of \cite{Gauntlett03} we read
\begin{equation}\label{eq04}
     S_{4E} = \int d^4x\, \sqrt{g_4}\, R^7 \left(- \left(\mathcal{R}_4-\frac{10}{3} \Lambda \right) + \frac{3}{4 R^4} H_{\mu \nu}\, H^{\mu \nu} + \frac{1}{12 R^2} H_{\mu \nu \rho}\, H^{\mu \nu \rho} \right),
\end{equation}
where $\mathcal{R}_4$ is the scalar curvature of $EAdS_4$ and $\Lambda = - \frac{12}{R^2}$ is the cosmological constant, and that with (\ref{eq03c}), we have
\begin{equation}\label{eq05}
     H_{\mu \nu}\, H^{\mu \nu} = \frac{32}{R^8}\, u^2 (\partial_i f_5)(\partial^i f_5), \quad H_{\mu \nu \rho}\, H^{\mu \nu \rho} = \frac{384}{R^{10}}\, u^2 (\partial_i f_5)(\partial^i f_5).
\end{equation}
As a result, the (pseudo)scalar equation from the action is (\ref{eq03d}) that in turn satisfies
\begin{equation}\label{eq06}
     d( R^2 \ast_4 H_3) + 6 \ast_4 H_2 =0,  \quad  d( R^3 \ast_4 H_2)=0,
\end{equation}
which are the (pseudo)scalar equations from (B.9)-(B.13) of \cite{Gauntlett03}.

On the other hand, from the 11D Einstein equation
\begin{equation}\label{eq07}
 \mathcal{R}_{MN}= \frac{1}{6} \left(\frac{3}{3!} {G}_{MPQR}\, {G}_N^{PQR} - \frac{1}{4!} g_{MN}\, {G}_{PQRS}\, {G}^{PQRS} \right),
\end{equation}
we read the equation
\begin{equation}\label{eq08}
(\partial_i f_5)(\partial^i f_5) + {C}_0\, \frac{R^{10}}{u^2} =0,
\end{equation}
with ${C}_0=\frac{2}{7}, \frac{1}{2}$ and $-\frac{2}{5}$ corresponding to $\mathcal{R}_{\mu \nu}$, $\mathcal{R}_{mn}$ and $\mathcal{R}_{77}$ from (B.19), (B.21) and (B.22) of \cite{Gauntlett03}, respectively. Then, a solution from (\ref{eq08}) is also realized as (\ref{eq03d}) with
\begin{equation}\label{eq09}
 c_4 = c_5\, \frac{r^2}{u},
\end{equation}
where $c_5$ has different values for different $C_0$'s. It is noteworthy that the solution is structurally similar to the (pseudo)scalar $m^2 =+4$ asymptotic solution near the boundary ($u = 0$), associated with the non-normalizable mode $\Delta_- = -1$ of course. It should also be noted that it is not possible to have a unique solution including the backreaction on the whole background geometry; In fact, to have just one solution (one $c_5$), we must consider the backreaction in one part of 11D space (e.g. the external space) and neglect it on the remaining parts.

\section{Details of Computing the Stress Tensors} \label{Appendix.A0}
Although we could discuss on the backreaction according to relations (B.19-22) of \cite{Gauntlett03} directly, we do it in our own way in this Appendix. For the Einstein equation
\begin{equation}\label{eq14a}
    \mathcal{R}_{MN} - \frac{1}{2} g_{MN} \mathcal{R} = 8 \pi \mathcal{G}_{11} T_{MN}^{\tilde{G}_4},
\end{equation}
we consider
\begin{equation}\label{eq14b}
 T_{MN}^{\tilde{G}_4} = \frac{1}{4!} \left[4 \tilde{G}_{MPQR} \tilde{G}_N^{PQR} - \frac{1}{2} g_{MN} \tilde{G}_{PQRS} \tilde{G}^{PQRS} \right],
\end{equation}
where we use the capital indices $M, N,...$ for the 11D space-time directions and small indices $m, n,...$ for the 6D internal $CP^3$ space, and the Greek indices $\mu, \nu,....$ for the 4D external $AdS_4$ space.

We compute the external and internal components of the above EM tensor for the ansatz (\ref{eq11}) wrt its dual 7-from, which in components read
\begin{equation}\label{eq15a}
 \tilde{G}_{PQRS} \equiv \hat{c}_1\ \tilde{G}_{m n p q} + \hat{c}_2\ \tilde{G}_{\mu m n 7} + \hat{c}_3\ \tilde{G}_{\mu \nu \rho \sigma},
\end{equation}
\begin{equation}\label{eq15b}
 \tilde{G}_{PQRSTUV} \equiv  \bar{f}_1\ \tilde{G}_{\mu \nu \rho \sigma m n 7} + \tilde{G}_{\mu \nu \rho m n p q 7} +  f_7\ \tilde{G}_{m n p q r s 7},
\end{equation}
respectively, where we use $e_7$ as the seventh vielbein (i.e. as a coordinate base) and that
\begin{equation}\label{eq16a}
 \begin{split}
 & \tilde{G}_{m n p q} = 6\, F_{m n p q},\quad \mathcal{F}_{m n p q} = J_{[{m n}} J_{{p q}]} = \frac{1}{3} \left(J_{m n}J_{p q} - J_{p n}J_{m q} - J_{q n}J_{p m}\right),  \\
 & \ \ \ \ \ \ \ \ \ \ \ \ \ \ \ \ \ \ \tilde{G}_{\mu m n 7} = (\partial_\mu f_3)\, J_{m n} = - \tilde{G}_{7 m n \mu}, \quad \tilde{G}_{\mu \nu \rho \sigma} = \varepsilon_{\mu \nu \rho \sigma}, \\
 & \ \ \ \ \ \ \ \ \ \ \ \ \ \ \ \ \ \hat{c}_1 = 8\, R^4\, \bar{f}_1, \quad  \hat{c}_2 = -2\, R^4 \quad  \hat{c}_3 = \frac{3}{8} R^4\, f_7,
 \end{split}
\end{equation}
\begin{equation}\label{eq16b}
 \begin{split}
\tilde{G}_{\mu \nu \rho \sigma m n 7} & = (105)\, \varepsilon_{\mu \nu \rho \sigma}\, J_{m n}, \quad \tilde{G}_{\mu \nu \rho m n p q 7} = (210)\, A_{\mu \nu \rho}\, \mathcal{F}_{m n p q}, \quad \tilde{G}_{m n p q r s 7} = \frac{7!}{8}\, \mathcal{F}_{m n p q r s}, \\
& \ \ \ \ \ \ \ \ \ \ \ \ \ \ \ \ \ \ \ \ \ \ \ \ \ \ \ \ \ \ \ A_{\mu \nu \rho}=\sqrt{g_4}\ g^{\sigma \sigma} \varepsilon_{\sigma \mu \nu \rho}\, \partial^{\sigma} f_3, \\
\mathcal{F}_{m n p q r s} & = J_{[{m n}} J_{p q} J_{{r s}]} = \frac{1}{15} (J_{m q} J_{n s} J_{p r} + J_{m s} J_{n p} J_{q r} - J_{m s} J_{n q} J_{p r} - J_{m r} J_{n p} J_{q s} + J_{m r} J_{n q} J_{p s} \\
  & + J_{m p} J_{n r} J_{q s}- J_{m p} J_{n s} J_{q r} - J_{m q} J_{n r} J_{p s} + J_{m n} J_{q r} J_{p s} + J_{n p} J_{m q} J_{r s} + J_{n r} J_{p q} J_{m s} \\
   & + J_{m n} J_{p q} J_{r s} - J_{m n} J_{q s} J_{p r} - J_{n q} J_{m p} J_{r s} - J_{n s} J_{p q} J_{m r} ).
 \end{split}
\end{equation}

To continue, we write
\begin{equation}\label{eq17a}
 \tilde{G}_{PQRS} \tilde{G}^{PQRS} = \hat{c}_1\, \tilde{G}_{m n p q} \tilde{G}^{m n p q} + 12\, \hat{c}_2\, \tilde{G}_{\mu m n 7} \tilde{G}^{\mu m n 7} + \hat{c}_3\, \tilde{G}_{\mu \nu \rho \sigma} \tilde{G}^{\mu \nu \rho \sigma},
\end{equation}
\begin{equation}\label{eq17b}
 \tilde{G}_{\mu PQR} \tilde{G}_\nu^{PQR} = 3\, \hat{c}_2\, \tilde{G}_{\mu m n 7} \tilde{G}_\nu^{m n 7} + \hat{c}_3\, \tilde{G}_{\mu \rho \sigma \delta} \tilde{G}_\nu^{\rho \sigma \delta},
\end{equation}
\begin{equation}\label{eq17c}
 \tilde{G}_{m PQR} \tilde{G}_n^{PQR} = \hat{c}_1\, \tilde{G}_{m p q r} \tilde{G}_n^{p q r} + 6\, \hat{c}_2\, \tilde{G}_{m p 7 \mu} \tilde{G}_n^{p 7 \mu},
\end{equation}
\begin{equation}\label{eq17d}
 \tilde{G}_{7 PQR} \tilde{G}_7^{PQR} = 3\, \hat{c}_2\ \tilde{G}_{7 m n \mu} \tilde{G}_7^{m n \mu},
\end{equation}
in which the numerical coefficients are for permutations of indices. As a result, we obtain
\begin{equation}\label{eq18a}
 \tilde{G}_{PQRS} \tilde{G}^{PQRS} = 96 \left[\frac{8}{R^7} \bar{f}_1^2 + \frac{1}{2 R^5} (\partial_\mu f_3)(\partial^\mu f_3) + \frac{3}{8 R^7} f_{7}^2\right],
\end{equation}
\begin{equation}\label{eq18b}
 \tilde{G}_{\mu PQR} \tilde{G}_\nu^{PQR} = \frac{9}{R^7} f_{7}^2\ g_{\mu \nu} + \frac{12}{R^5} (\partial_\mu f_3)(\partial_\nu f_3),
\end{equation}
\begin{equation}\label{eq18c}
 \tilde{G}_{m PQR} \tilde{G}_n^{PQR} = \frac{128}{R^7} \bar{f}_1^2\ g_{m n} + \frac{4}{R^5} (\partial_\mu f_3)(\partial^\mu f_3)\ g_{m n} ,
\end{equation}
\begin{equation}\label{eq18d}
 \tilde{G}_{7 PQR} \tilde{G}_7^{PQR} = \frac{12}{R^5} (\partial_\mu f_3)(\partial^\mu f_3)\ g_{7 7},
\end{equation}
with a $4!$ coefficient for all terms. In obtaining the latter results, we have used the differential-geometry formulas in Appendix \ref{Appendix.A1} together with permutations of the indices depended on their locations on the forms besides the following relations for the K\"{a}hler form on $CP^3$:
\begin{equation}\label{eq19}
 \begin{split}
  & \ \ \ \ J_{m n} = - J_{n m}, \quad J_{m n} J^{m n} = 6, \quad g_{m p} g_n^{\ p} = g_{m n}, \\
  & J_{m n} J_{p q} J_{r s}\ \varepsilon^{\acute{m} n p q r s} = 8 g_m^{\ \acute{m}}, \quad J_{m n} J_{p q} J_{r s}\ \varepsilon^{m n p q r s} =48.
 \end{split}
\end{equation}

Then, plugging (\ref{eq18a}) with (\ref{eq18b}), (\ref{eq18c}) and (\ref{eq18d}) back into (\ref{eq14b}), making use of (\ref{eq13b}), taking traces and using the Euler-Lagrange equation, we finally arrive at
\begin{equation}\label{eq20a}
     \Box_4 \bar{f}_1 - \left(2 \pm 6\, \bar{C}_7 \right) \frac{\bar{f}_1}{R^2} - 192\, \bar{f}_1^3=0,
\end{equation}
\begin{equation}\label{eq20b}
     \Box_4 \bar{f}_1 - \left(1 \pm 9\, \bar{C}_7 \right) \frac{\bar{f}_1}{R^2} - 288\, \bar{f}_1^3=0,
\end{equation}
\begin{equation}\label{eq20c}
     \Box_4 \bar{f}_1 - \left(-1 \pm 3\, \bar{C}_7 \right) \frac{\bar{f}_1}{R^2} - 96\, \bar{f}_1^3=0,
\end{equation}
respectively.

\subsection{Some Useful Formulas}\label{Appendix.A1}
The main differential-geometry relations, in order to do the computations of Appendix \ref{Appendix.A0}, in general D-dimensions read
\begin{equation}\label{eq21a}
 \tilde{G}_\alpha = \frac{1}{\alpha!} \tilde{G}_{R_1 R_2 ... R_\alpha} dX^{R_1 R_2 ... R_\alpha}, \quad dX^{R_1 R_2 ... R_\alpha} \equiv dX^{R_1} \wedge dX^{R_2} \wedge ... \wedge dX^{R_\alpha},
\end{equation}
\begin{equation}\label{eq21b}
 \ast_D \tilde{G}_\alpha = \frac{\sqrt{g_D}}{(D- \alpha)!\ \alpha!} \varepsilon_{R_1 R_2 ... R_{\alpha} R_{\alpha +1} ... R_{D- \alpha}} \tilde{G}^{R_1 R_2 ... R_\alpha} dX^{R_{\alpha +1} ... R_{D- \alpha}},
\end{equation}
\begin{equation}\label{eq21c}
 \tilde{G}_\alpha \wedge \tilde{H}_\beta = \frac{1}{\alpha!} \frac{1}{\beta!} \tilde{G}_{R_1 R_2 ... R_\alpha} \tilde{H}_{S_1 S_2 ... S_\beta}\ dX^{R_1 R_2 ... R_\alpha S_1 S_2 ... S_\beta},
\end{equation}
\begin{equation}\label{eq21d}
 \tilde{G}_\alpha \wedge \ast_D \tilde{G}_\alpha = \frac{\sqrt{g_D}}{\alpha!} \tilde{G}_{R_1 R_2 ... R_D} \tilde{G}^{R_1 R_2 ... R_D}\ dX^{1 2 ... D},
\end{equation}
\begin{equation}\label{eq21e}
 dX^{1 2 ... D} = \frac{1}{D!}\varepsilon_{R_1 R_2 ... R_D} dX^{R_1 R_2 ... R_D},
\end{equation}
\begin{equation}\label{eq21f}
 \varepsilon^{R_1 R_2 ... R_\alpha S_1 S_2 ... S_{D-\alpha}}\  \varepsilon_{R_1 R_2 ... R_\alpha T_1 T_2 ... T_{D-\alpha}} = \alpha! (D-\alpha)!\ \delta^{[S_1...}_{T_1 ...} \delta^{S_{D-\alpha}]}_{T_{D-\alpha}},
\end{equation}
for the $\alpha$-form $\tilde{G}$ and $\beta$-form $\tilde{H}$. It is also notable that we have used
\begin{equation}\label{eq22}
 \tilde{G}^{PQRS} = \frac{1}{\sqrt{g_{11}}\ 7!}\, \varepsilon^{PQRS R_1 R_2 ... R_7} \tilde{G}_{R_1 R_2 ... R_7},
\end{equation}
as a result of (\ref{eq21b}) with $D=11$.

\end{appendices}


\begin{thebibliography}{99}
\bibitem{Aharony99} O. Aharony, S. S. Gubser, J. Maldacena, H. Ooguri and Y. Oz, \textit{"Large $N$ field theories, string theory and gravity"}, Phys. Rept. 323, 183 (2000), \href{http://arxiv.org/abs/hep-th/9905111}{[arXiv:hep-th/9905111]}.
\bibitem{Me7} M. Naghdi, \textit{"Massive (pesudo)Scalars in AdS$_4$, SO(4) Invariant Solutions and Holography"}, \href{http://arxiv.org/abs/1703.02765}{[arXiv:1703.02765 [hep-th]]}.
\bibitem{Gauntlett03} J. P. Gauntlett, S. Kim, O. Varela and D. Waldram, \textit{"Consistent supersymmetric Kaluza--Klein
    truncations with massive modes"}, JHEP 0904, 102 (2009), \href{http://arxiv.org/abs/0901.0676}{[arXiv:0901.0676 [hep-th]]}.
\bibitem{Me4} M. Naghdi, \textit{"Marginal fluctuations as instantons on M2/D2-branes"}, Eur. Phys. J. C 74, 2826 (2014),
    \href{http://arxiv.org/abs/1302.5534}{[arXiv:1302.5534 [hep-th]]}.
\bibitem{Bergshoeff96} E. Bergshoeff, M. de Roo, E. Eyras, B. Janssen and J. P. van der Schaar, \textit{"Multiple intersections
of D-branes and M-branes"}, Nucl. Phys. B 494, 119 (1997), \href{http://arxiv.org/abs/hep-th/9612095}{[arXiv:hep-th/9612095]}.
\bibitem{Elitzur05} S. Elitzur, A. Giveon, M. Porrati and E. Rabinovici, \textit{"Multitrace deformations of vector and adjoint theories and their holographic duals"}, JHEP 0602, 006 (2006), \href{http://arxiv.org/abs/hep-th/0511061}{[arXiv:hep-th/0511061]}.
\bibitem{ABJM} O. Aharony, O. Bergman, D. L. Jafferis and J. Maldacena, \textit{"$\mathcal{N}$=6 superconformal Chern-Simons
    matter theories, M2-branes and their gravity duals"}, JHEP 0810, 091 (2008), \href{http://arxiv.org/abs/0806.1218}{[arXiv:0806.1218[hep-th]]}.
\bibitem{CrapsHertog} B. Craps, T. Hertog and N. Turok, \textit{"A multitrace deformation of ABJM theory"}, Phys. Rev. D 80, 086007 (2009), \href{http://arxiv.org/abs/0905.0709}{[arXiv:0905.0709 [hep-th]]}.
\bibitem{Gauntlett0912} J. Gauntlett, J. Sonner and T. Wiseman, \textit{"Quantum criticality and holographic superconductors in M-theory"}, JHEP 1002, 060 (2010), \href{http://arxiv.org/abs/0912.0512}{[arXiv:0912.0512 [hep-th]]}.
\bibitem{Me6} M. Naghdi, \textit{"Non-minimally coupled pseudoscalars in $AdS_4$ for instantons in CFT$_3$"}, Class. Quant. Grav. 33, 115005 (2016), \href{http://arxiv.org/abs/1505.00179}{[arXiv:1505.00179 [hep-th]]}.
\bibitem{FreedmanNicolai} D. Z. Freedman and H. Nicolai, \textit{"Multiplet shortening in Osp(N,4)"}, \href{http://www.sciencedirect.com/science/article/pii/0550321384901640} {Nucl. Phys. B 237, 342 (1984)}.
\bibitem{NilssonPope} B. E. W. Nilsson and C. N. Pope, \textit{"Hopf fibration of eleven-dimensional supergravity"},
    \href{http://iopscience.iop.org/0264-9381/1/5/005/}{Class. Quant. Grav. 1, 499 (1984)}.
\bibitem{Biran} B. Biran, A. Casher, F. Englert, M. Rooman and P. Spindel, \textit{"The fluctuating seven-sphere in eleven-dimensional supergravity"}, \href{http://www.sciencedirect.com/science/article/pii/037026938490666X} {Phys. Lett. B 134, 179 (1984)}.
\bibitem{DuffNilssonPope84} M. J. Duff, B. E. W. Nilsson and C. N. Pope, \textit{"Superunification from eleven dimensions"}, \href{http://www.sciencedirect.com/science/article/pii/0550321384905777?np=y} {Nucl. Phys. B 233, 433 (1984)}.
\bibitem{N} M. Naghdi, \textit{"A monopole Instanton-like effect in the ABJM model"}, Int. J. Mod. Phys. A 26, 3259 (2011),
    \href{http://arxiv.org/abs/1106.0907}{[arXiv:1106.0907 [hep-th]]}.
\bibitem{Me5} M. Naghdi, \textit{"Dual localized objects from M-branes over $AdS_4 \times S^7/Z_k$"},  Class. Quant. Grav. 32, 215018 (2015), \href{http://arxiv.org/abs/1502.03281}{[arXiv:1502.03281 [hep-th]]}.
\bibitem{Me3} M. Naghdi, \textit{"New instantons in AdS$_4$/CFT$_3$ from D4-branes wrapping some of CP$^3$"}, Phys. Rev. D 88, 026013 (2013), \href{http://arxiv.org/abs/1302.5294}{[arXiv:1302.5294 [hep-th]]}.
\bibitem{ABJ} O. Aharony, O. Bergman, D. L. Jafferis, \textit{"Fractional M2-branes"}, JHEP 0811, 043 (2008), \href{http://arxiv.org/abs/0807.4924}{[arXiv:0807.4924 [hep-th]]}.
\bibitem{Giombi01} S. Giombi and X. Yin, \textit{"The higher spin/vector model duality"}, J. Phys. A 46, 214003 (2013), \href{http://arxiv.org/abs/1208.4036}{[arXiv:1208.4036 [hep-th]]}.
\bibitem{Aharony1110} O. Aharony, G. Gur-Ari and R. Yacoby, \textit{"$d=3$ bosonic vector models coupled to Chern-Simons gauge theories"}, JHEP 1203, 037 (2012), \href{http://arxiv.org/abs/1110.4382}{[arXiv:1110.4382 [hep-th]]}.
\bibitem{Minwalla1207} C.-Ming Chang, Sh. Minwalla, T. Sharma and X. Yin, \textit{"ABJ triality: from higher spin fields to strings"}, J. Phys. A 46, 214009 (2013), \href{http://arxiv.org/abs/1207.4485}{[arXiv:1207.4485 [hep-th]]}. \href{http://arxiv.org/abs/hep-th/0511061}{[arXiv:hep-th/0511061]}.
\bibitem{deharo06} S. de Haro and A. C. Petkou, \textit{"Instantons and conformal holography"}, JHEP 0612, 076 (2006), \href{http://arxiv.org/abs/hep-th/0606276}{[arXiv:hep-th/0606276]}.
\bibitem{deHaro2} S. de Haro, I. Papadimitriou and A. C. Petkou, \textit{"Conformally coupled scalars, instantons and vacuum instability in $AdS_4$"}, Phys. Rev. Lett. 98, 231601 (2007), \href{http://arxiv.org/abs/hep-th/0611315}{[arXiv:hep-th/0611315]}.
\bibitem{Barbon1003} J. L. F. Barbon and E. Rabinovici, \textit{"Holography of AdS vacuum bubbles"}, JHEP 1004, 123 (2010), \href{http://arxiv.org/abs/1003.4966}{[arXiv:1003.4966 [hep-th]]}.
\bibitem{Maldacena010} J. Maldacena, \textit{"Vacuum decay into Anti de Sitter space"}, \href{http://arxiv.org/abs/1012.0274}{[arXiv:1012.0274 [hep-th]]}.
\bibitem{Bena} I. Bena, \textit{"The M theory dual of a three-dimensional theory with reduced supersymmetry"}, Phys. Rev. D 62, 126006 (2000), \href{http://arxiv.org/abs/hep-th/000414}{[arXiv:hep-th/000414]}.
\bibitem{BzowskiHertog} A. Bzowski, T. Hertog and M. Schillo, \textit{"Cosmological singularities encoded in IR boundary correlations"}, JHEP 1605, 168 (2016), \href{http://arxiv.org/abs/1512.05761}{[arXiv:1512.05761 [hep-th]]}.
\bibitem{SmolkinTurok} M. Smolkin and N. Turok, \textit{"Dual description of a 4d cosmology"},
    \href{http://arxiv.org/abs/1211.1322}{[arXiv:1211.1322 [hep-th]]}.
\bibitem{Gomiz} J. Gomis, D. Rodriguez-Gomez, M. Van Raamsdonk and H. Verlinde, \textit{"A massive study of M2-brane proposals"}, JHEP 0809, 113 (2008), \href{http://arxiv.org/abs/0807.1074}{[arXiv:0807.1074 [hep-th]]}.
\bibitem{Terashima} S. Terashima, \textit{"On M5-branes in $\mathcal{N}=6$ membrane action"}, JHEP 0808, 080 (2008), \href{http://arxiv.org/abs/0807.0197}{[arXiv:0807.0197 [hep-th]]}.
\bibitem{Fubini1} S. Fubini, \textit{"A new approach to conformal invariant field theories"},
     \href{http://link.springer.com/article/10.1007%2FBF02785664}{Nuovo Cim. A 34, 521 (1976)}.

\bibitem{HanakiLin} K. Hanaki and H. Lin, \textit{"M2-M5 systems in $\mathcal{N}$=6 Chern-Simons theory"}, JHEP 0809, 067 (2008), \href{http://arxiv.org/abs/0807.2074}{[arXiv:0807.2074 [hep-th]]}.
\bibitem{Bardeen1984} W. A. Bardeen, M. Moshe and M. Bander, \textit{"Spontaneous breaking of scale invariance and the ultraviolet fixed point in $O(N)$-Symmetric ($\varphi_3^6$) Theory"}, \href{http://journals.aps.org/prl/abstract/10.1103/PhysRevLett.52.1188}{Phys. Rev. Lett. 52, 1188 (1984)}.
\bibitem{RabinoviciSmolkin011} E. Rabinovici and M. Smolkin, \textit{"On the dynamical generation of the Maxwell term and scale invariance"}, JHEP 1107, 040 (2011), \href{http://arxiv.org/abs/1102.5035}{[arXiv:1102.5035 [hep-th]]}.
\bibitem{BardeenMoshe014} W. A. Bardeen and M. Moshe, \textit{"Spontaneous breaking of scale invariance in a $D=3\ U(N)$ model with Chern-Simons gauge field"}, JHEP 1406, 113 (2014), \href{http://arxiv.org/abs/1402.4196}{[arXiv:1402.4196 [hep-th]]}.
\bibitem{Vasiliev01} M. A. Vasiliev, \textit{"Higher spin gauge theories: Star product and AdS space"}, In "the many faces of the superworld: pp. 533-610", \href{http://arxiv.org/abs/hep-th/9910096}{[arXiv:hep-th/9910096]}.
\bibitem{HertogMaeda01} T. Hertog and K. Maeda, \textit{"Black holes with scalar hair and asymptotics in N=8 supergravity"},
    JHEP 0407, 051 (2004), \href{http://arxiv.org/abs/hep-th/0404261}{[arXiv:hep-th/0404261]}.
\bibitem{BarbonRabinovici011} J. L. F. Barbon and E. Rabinovici, \textit{"AdS crunches, CFT falls and cosmological complementarity"}, JHEP 1104, 044 (2011), \href{http://arxiv.org/abs/1102.3015}{[arXiv:1102.3015 [hep-th]]}.
\bibitem{BumHoonLee014A} B.-H. Lee, W. Lee, D. Ro and D.-h. Yeom, \textit{"Generalized Fubini instantons of a self-gravitating scalar field"}, \href{https://link.springer.com/article/10.3938%2Fjkps.65.884}{J. Korean Phys. Soc. 65, 884 (2014)}.

\bibitem{BumHoonLee014} B.-H. Lee, W. Lee, D. Ro and D.-h. Yeom, \textit{"Oscillating Fubini instantons in curved space"}, Phys. Rev. D 91, 124044 (2015), \href{http://arxiv.org/abs/1409.3935}{[arXiv:1409.3935 [hep-th]]}.

\end{thebibliography}
\end{document}